\documentclass[aps,twocolumn,nobibnotes,nofootinbib,%
superscriptaddress,noshowpacs,preprintnumbers]{revtex4}

\usepackage{graphicx}

\begin{document}

\title{Mimicking diffuse supernova antineutrinos
with the Sun as a source\footnote{Contribution to a special issue of
Yadernaya Fizika (Physics of Atomic Nuclei) on occasion of Lev
Borisovich Okun's 80th birthday.}}

\author{\firstname{Georg} \surname{Raffelt}}
\affiliation{Max-Planck-Institut f\"ur Physik
(Werner-Heisenberg-Institut) F\"ohringer Ring 6, 80805 M\"unchen,
Germany}

\author{\firstname{Timur} \surname{Rashba}}
\affiliation{Max-Planck-Institut f\"ur Sonnensystemforschung,
Max-Planck-Str.~2, 37191 Katlenburg-Lindau, Germany}
\affiliation{Institute of Terrestrial Magnetism, Ionosphere and
Radio Wave Propagation of the Russian Academy of Sciences, Troitsk,
Moscow region, 142190, Russia}

\begin{abstract}
Measuring the $\bar\nu_e$ component of the cosmic diffuse supernova
neutrino background (DSNB) is the next ambitious goal for low-energy
neutrino astronomy. The largest flux is expected in the lowest
accessible energy bin. However, for $E\alt15$~MeV a possible signal
can be mimicked by a solar $\bar\nu_e$ flux that originates from the
usual ${}^8$B neutrinos by spin-flavor oscillations. We show that such
an interpretation is possible within the allowed range of neutrino
electromagnetic transition moments and solar turbulent field strengths
and distributions. Therefore, an unambiguous detection of the DSNB
requires a significant number of events at $E\agt15$~MeV.
\end{abstract}

\preprint{MPP-2009-28}

\maketitle

\section{Introduction}                         \label{sec:introduction}

Neutrino astronomy and oscillation physics both began with the
pioneering Homestake observations of solar
neutrinos~\cite{Cleveland:1998nv}. For a long time the now common
interpretation of the solar neutrino deficit in terms of flavor
oscillations was not unique and indeed the apparent time variation
of the early data suggested magnetic spin or spin-flavor
oscillations as an intriguing possibility in which Lev Okun took a
keen interest~\cite{Okun:1986hi, Okun:1986na}. It was only the
KamLAND measurements of reactor neutrino
oscillations~\cite{Eguchi:2002dm} that proved beyond doubt that such
effects could not play a dominant role.

Today the frontiers of neutrino physics have shifted. The upcoming
generation of flavor oscillation experiments involves long-baseline
laboratory setups using reactors or high-energy beams as sources
whereas one frontier of neutrino astronomy relies on high-energy
neutrino telescopes. The new low-energy frontier includes solar
neutrino spectroscopy and the search for supernova (SN)
neutrinos~\cite{Autiero:2007zj}. While a high-statistics neutrino
light curve from the next galactic SN would be a dream come true, in
the meantime one may plausibly search for the cosmic diffuse SN
neutrino background (DSNB) that is provided by all past
SNe~\cite{Guseinov, Hartmann:1997qe, Ando:2004hc, Lunardini:2005jf,
Horiuchi:2008jz, Lunardini:2009ya}.  In the energy range $10~{\rm
MeV}\alt E\alt30~{\rm MeV}$ the DSNB, with a total flux of order
$10~{\rm cm}^{-2}~{\rm s}^{-1}$, exceeds the atmospheric neutrino
flux and other backgrounds and could provide interesting information
on the stellar core-collapse rate and the corresponding neutrino
spectra.

Cosmic redshift implies that the expected DSNB peaks at low energies
and the largest number of events is expected in the lowest
accessible energy bin. The only realistic detection process is
inverse beta decay, $\bar\nu_e+p\to n+e^+$, so reactor $\bar\nu_e$
fluxes are an unsurmountable background below about 10~MeV.

We here study another background that can be important below about
15~MeV, the $\nu_e\to\bar\nu_e$ conversion of solar
neutrinos~\cite{Hartmann:1997qe}. In particular, we consider
spin-flavor conversions that are induced by solar magnetic fields if
neutrinos are Majorana particles and have sizeable electric or
magnetic transition moments~\cite{Schechter:1981hw, Akhmedov:1988uk,
Lim:1987tk, Miranda:2003yh, Miranda:2004nz, Friedland:2005xh}. While
the old spin-flavor interpretation of the solar neutrino deficit
would have required a conversion rate of order unity, here a
subdominant effect of order $10^{-6}$ is enough to mimic the DSNB
for $E\alt15$~MeV. Such a conversion rate can be achieved without
extreme assumptions about solar magnetic fields and within the
allowed range of neutrino electromagnetic transition moments. The
latter, of course, must be much larger than suggested by
standard-model physics. If one takes this possibility seriously, the
first events observed in the low-energy range can not be uniquely
attributed to the DSNB unless there is enough statistics to
reconstruct reliable spectral information.

To illustrate our point we begin in Sec.~\ref{sec:dsnb} with a brief
review of the DSNB and its detection possibilities.  In
Sec.~\ref{sec:sun} we turn to the Sun as a source of antineutrinos
due to the spin-flavor conversion mechanism.  We discuss our
findings in Sec.~\ref{sec:discussion}.

\section{The diffuse supernova neutrino background}    \label{sec:dsnb}

The $\bar\nu_e$ component of the diffuse supernova neutrino background
(DSNB) can be expressed as~\cite{Ando:2004hc}
\begin{equation}
\label{eq:dsn}
\frac{d\phi_{\bar\nu_e}^{\rm DSNB}}{dE}=
\frac{1}{H_0}
\int\limits_0^{\infty}
\frac{dN(E_z)}{dE_z}R_{\rm SN}(z)
\frac{dz}{\sqrt{(z+1)^3\Omega_{\rm M}+\Omega_\Lambda}}\,,
\end{equation}
where $E_z=(1+z)\,E$ is the neutrino energy at redshift $z$, $E$ is
the present-day neutrino energy, $N(E_z)$ is the $\bar\nu_e$
spectrum of an individual SN, $R_{\rm SN}(z)$ the cosmic SN
rate at redshift $z$, and
$H_0=73\,\mbox{km}~\mbox{s}^{-1}~\mbox{Mpc}^{-1}$ the Hubble
constant. $\Omega_{\rm M}=0.27$ and $\Omega_\Lambda=0.73$ are the matter
and dark energy density, respectively~\cite{Amsler:2008zzb}.

To be specific we parametrize the cosmic SN rate in the
form~\cite{Ando:2004hc}
\begin{eqnarray}
R_{\rm SN}(z)&=&4.1\times10^{-3}\,\mbox{yr}^{-1}~\mbox{Mpc}^{-3}
\,f_{\rm SN}\,h_{73}\nonumber\\*
&\times&\frac{e^{3.4z}}{e^{3.8z}+45}
\left[\Omega_{\rm M}+\frac{\Omega_\Lambda}{(z+1)^{3}}\right]^{1/2}\,,
\end{eqnarray}
where $f_{\rm SN}$ is a normalization factor of order of unity and
$h_{73}$ is $H_0$ in units of
$73\,\mbox{km}~\mbox{s}^{-1}~\mbox{Mpc}^{-1}$.

The average $\bar\nu_e$ spectrum emitted by a SN is expressed in the
quasi-thermal form~\cite{Keil:2002in}
\begin{equation}
\frac{dN(E)}{dE}=\frac{(1+\alpha)^{1+\alpha}E_{\rm tot}}
{\Gamma(1+\alpha)\bar E^2}
\left(\frac{E}{\bar E}\right)^{\alpha}e^{-(1+\alpha)E/\bar E}\,,
\end{equation}
where we use $\bar E=15$~MeV for the average energy, $\alpha=4$ for
the pinching parameter, and $E_{\rm tot}=5\times10^{52}$~erg for the
total amount of energy emitted in $\bar\nu_e$. The emitted spectrum
is understood after all oscillation effects. The flavor-dependent
differences of the antineutrino spectra emitted at the neutrino
sphere are not large~\cite{Keil:2002in}, so oscillation effects are
not a major concern.

\begin{figure}[b]
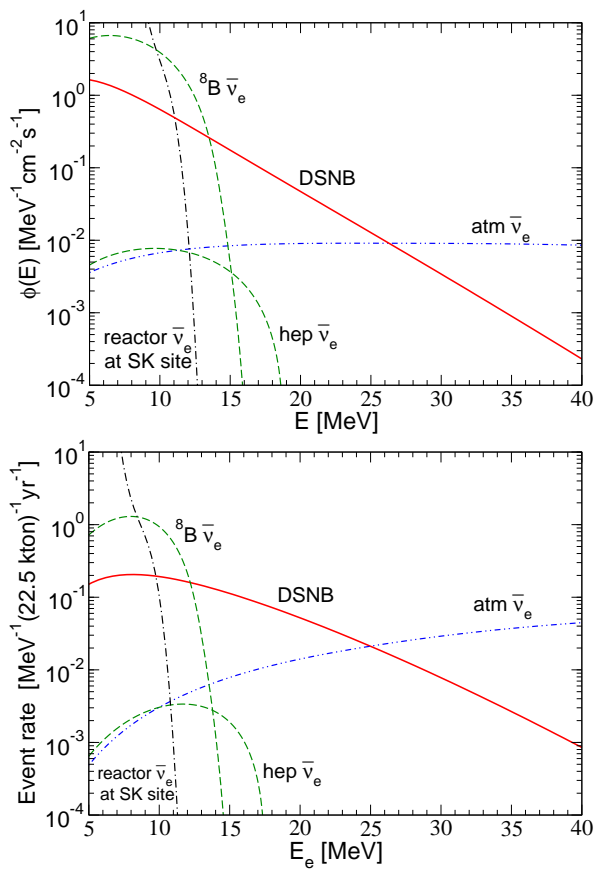

\includegraphics[width=.9\columnwidth]{fig1.eps}
\includegraphics[width=.9\columnwidth]{fig2.eps}
\caption{$\bar\nu_e$ fluxes at the Super-Kamiokande site (upper
  panel) and positron event spectra (lower panel).
  The DSNB is an estimate for typical parameters (see
  text). The solar $\bar\nu_e$ fluxes correspond to the effective
  $\nu_e\to\bar\nu_e$ conversion probability of Eq.~(\ref{eq:flux}),
  i.e.\ $10^{-5}$ at $E=10$~MeV. \label{fig:dsn}}
\end{figure}

The $\bar\nu_e$ component of the DSNB calculated with these
assumptions serves as our benchmark case and is shown in
Fig.~\ref{fig:dsn}. The uncertainty in amplitude and spectral shape
are considerable and we refer to the literature for a
discussion~\cite{Ando:2004hc, Lunardini:2005jf, Horiuchi:2008jz,
Lunardini:2009ya}.

At low energies, the DSNB is overwhelmed by the background
$\bar\nu_e$ flux from reactors. In Fig.~\ref{fig:dsn} we show an
estimate for the Super-Kamiokande site using an approximate
analytical expression~\cite{Clemens}. This flux significantly
exceeds those expected at other possible locations that have been
discussed, for example, by the LENA
collaboration~\cite{Wurm:2007cy}.
At high energies, the limiting factor is the atmospheric neutrino
flux. The estimate shown in Fig.~\ref{fig:dsn} is based on
Ref.~\cite{Gaisser:1988ar}.

The only realistic reaction for detecting the DSNB is inverse beta
decay $\bar\nu_e+p\to n+e^+$. Therefore, we show in the lower panel
of Fig.~\ref{fig:dsn} the same fluxes modulated with the cross section
of this reaction, i.e., the expected event spectra. The largest number
of events is expected in the lowest useful energy bin above the
reactor background.

The most stringent upper limit is $1.08\,{\rm cm}^{-2}~{\rm s}^{-1}$
at 90\% CL for $E>19.3$~MeV, obtained by the Super-Kamiokande
experiment \cite{Malek:2002ns}. An actual detection requires tagging
the final-state neutrons. In a large future scintillator detector
such as the proposed LENA~\cite{Wurm:2007cy} neutron tagging is part
of the detection signature. In water Cherenkov detectors, neutron
tagging requires gadolinium loading~\cite{Beacom:2003nk}, a
possibility that is being investigated for Super-Kamiokande. Both
detector types probably can lower the energy threshold all the way
to the reactor background.

\section{Solar Antineutrinos}                           \label{sec:sun}

If the detection threshold indeed can be lowered to the energy where
the reactor background begins to dominate, the first DSNB events would
be expected in the lowest energy bin above this boundary. In this
range the solar $\nu_e$ flux from the ${}^8$B reaction is more than
five orders of magnitude above the baseline DSNB flux shown in
Fig.~\ref{fig:dsn}. Therefore, even a $\nu_e\to\bar\nu_e$ conversion
efficiency as small as $10^{-6}$ is enough to provide a significant
background.

If neutrinos are Majorana particles and have non-zero transition
magnetic moments, then solar electron neutrinos can oscillate to
antineutrinos in the solar magnetic field~\cite{Schechter:1981hw,
  Akhmedov:1988uk, Lim:1987tk}. Little is known about the interior
$B$-field distribution in the Sun.  We assume turbulent fields in the
convective zone because they are well motivated and lead to the
strongest conversion effect~\cite{Miranda:2003yh, Miranda:2004nz}.

For turbulent fields, a simple analytic expression for the
$\nu_e\to\bar\nu_e$ conversion probability is~\cite{Miranda:2003yh,
Miranda:2004nz, Friedland:2005xh}
\begin{eqnarray}
\label{eq:flux}
P&\approx&
10^{-5}S^2 \mu_{11}^2
\left(\frac{B}{20~\mbox{kG}}\right)^2
\left(\frac{3\times10^4~\mbox{km}}{L_{\rm max}}\right)^{p-1}
\nonumber\\*
&\times&
\left(\frac{8\times10^{-5}\,\mbox{eV}^2}{\Delta
m_\odot^2}\right)^{p} \left(\frac{E}{10\,\mbox{MeV}}\right)^{p}
\left(\frac{\cos^2\theta_\odot}{0.7}\right).
\end{eqnarray}
Here $\mu_{11}=\mu_\nu/10^{-11}\mu_{\rm B}$ (Bohr magneton $\mu_{\rm
  B}=e/2m_e$) is the neutrino transition moment in a two-flavor
scenario, whereas $\Delta m_\odot^2$ and $\cos^2\theta_\odot$ are the
solar neutrino mixing parameters. $S$ is a factor of order unity
describing the spatial configuration of the magnetic field, $B$ is the
average strength of the magnetic field at spatial scale $L_{\rm max}$
(the size of the largest eddies at which energy is pumped to generate
turbulent motion), and $p$ is the power of the turbulence scaling. A
typical case is $p=\frac{5}{3}$ (Kolmogorov turbulence) whereas
conservative values for the other field parameters are $B=20$~kG and
$L_{\rm max}=3\times10^4$~km \cite{Miranda:2004nz, Friedland:2005xh}.

Neutrinos with nonvanishing masses and mixings inevitably have
nonvanishing electric and/or magnetic transition
moments~\cite{Pal:1981rm}, which are however proportional to the
neutrino masses and therefore extremely small.  The best experimental
limit on a neutrino transition moment connected to $\nu_e$ was found
by the Borexino collaboration to be $\mu_\nu<5.4 \times
10^{-11}\,\mu_{\rm B}$ at 90\% CL \cite{Arpesella:2008mt}, while the
best reactor limit is $\mu_\nu<5.8 \times 10^{-11}\,\mu_{\rm B}$ at
90\% CL obtained by the GEMMA experiment~\cite{Beda:2007hf}. An
astrophysical constraint to avoid excessive energy losses by
globular-cluster stars is $\mu_\nu<3 \times 10^{-12}\mu_{\rm
  B}$~\cite{Raffelt:1990pj, Raffelt:1992pi}.

Using the benchmark values for all parameters as in
Eq.~(\ref{eq:flux}) we show the expected solar $\bar\nu_e$ flux in
Fig.~\ref{fig:dsn}. It is much smaller than the direct KamLAND limit
on a possible solar $\bar\nu_e$ flux~\cite{Eguchi:2003gg}. In the
energy range above the reactor background and below the upper end of
the solar ${}^8$B spectrum, the solar $\bar\nu_e$ flux exceeds the
DSNB by about an order of magnitude. A $\mu_\nu$ on the level of the
globular-cluster limit still provides a flux comparable to the
DSNB. The true DSNB can be smaller than our benchmark by perhaps a
factor of ten whereas the solar $B$-field parameters can be more
favorable for spin-flavor oscillations. Therefore, it is clear that a
first $\bar\nu_e$ detection in this energy bin can be caused by the
Sun as a source instead of the DSNB.

In this case spin-flavor oscillations would also operate in the SN
environment and thus affect the DSNB in that the neutrino and
antineutrino source spectra would be partially swapped.  This effect
introduces an additional uncertainty in the DSNB prediction.

\section{Discussion}                             \label{sec:discussion}

Measuring the expected DSNB is a considerable challenge even with
the next generation of low-energy $\bar\nu_e$ detectors such as a
Gd-loaded version of Super-Kamiokande or a large-scale scintillator
detector like the proposed LENA. Our benchmark DSNB shown in
Fig.~\ref{fig:dsn} provides only a few events per year in a
Super-Kamiokande sized detector, and the true flux can be even
smaller. The expected spectrum decreases quickly with energy, so the
largest event rate is expected in the lowest accessible energy bin,
the reactor $\bar\nu_e$ background providing a hard lower boundary
near 10~MeV.

Therefore, it is tempting to focus on the energy range directly
above the reactor background. In this range, up to about 15~MeV, the
solar $\nu_e$ flux is huge, about six orders of magnitude larger
than the DSNB. Therefore, a small $\nu_e\to\bar\nu_e$ conversion
rate is enough to mimic the DSNB in this energy bin. While such a
conversion process violates lepton number, neutrinos with mass are
usually thought to be Majorana particles and lepton-number violation
is naturally present. A conversion probability on the $10^{-6}$ level
is naturally found in the framework of spin-flavor oscillations if
one assumes the presence of neutrino electric or magnetic transition
moments on the level of existing limits and middle-of-the-road
assumptions about solar turbulent $B$-field distributions.

Detecting a signature for lepton-number violation in the form of
solar $\nu_e\to\bar\nu_e$ conversions arguably would be a more
fundamental discovery than the DSNB itself. In this sense our
arguments can be turned around and the DSNB can be viewed as a
background to such a search.

Either way, disentangling the DSNB and a signature of solar
$\nu_e\to\bar\nu_e$ conversion is extremely difficult except by
enough statistics to provide spectral information above the endpoint
of the solar neutrino spectrum. Directional information based on the
neutron forward displacement in inverse $\beta$ decay requires even
larger event rates~\cite{Apollonio:1999jg, Hochmuth:2007gv}.

The uncertainties of the interior solar $B$-fields are large and
significant improvements on neutrino magnetic moment limits are not
foreseeable. Therefore, a clear detection of the DSNB likely will have
to depend on the energy range above the solar neutrino spectrum.

\section*{Acknowledgments}

TR thanks Cecilia Lunardini, Omar Miranda and Stan Woosley for
valuable and stimulating discussions.  TR was supported by a Marie
Curie International Incoming Fellowship of the European Community,
Contract No.~MIF1-CT-2005-008553. GR acknowledges partial support by
the Deutsche Forschungsgemeinschaft (DFG) under grant TR27
``Neutrinos and Beyond'' and the Cluster of Excellence ``Origin and
Structure of the Universe.''

\end{document}